# VICS82: THE VISTA–CFHT STRIPE 82 NEAR-INFRARED SURVEY

J. E. GEACH[1], Y.-T. LIN[2], M. MAKLER[3], J.-P. KNEIB[4,5], N. P. ROSS[6], W.-H. WANG[2], B.-C. HSIEH[2], A. LEAUTHAUD[7], K. BUNDY[7], H. J. MCCRACKEN[8], J. COMPARAT[9], G. B. CAMINHA[10], P. HUDELOT[8], L. LIN[2], L. VAN WAERBEKE[11], M. E. S. PEREIRA[3], AND D. MAST[3,12]

## ABSTRACT

We present the VISTA–CFHT Stripe 82 (VICS82) survey: a near-infrared ($J+K_s$) survey covering 150 square degrees of the Sloan Digital Sky Survey (SDSS) equatorial Stripe 82 to an average depth of $J$ = 21.9 AB mag and $K_s$ = 21.4 AB mag (80% completeness limits; $5\sigma$ point source depths are approximately 0.5 mag brighter). VICS82 contributes to the growing legacy of multi-wavelength data in the Stripe 82 footprint. The addition of near-infrared photometry to the existing SDSS Stripe 82 coadd *ugriz* photometry reduces the scatter in stellar mass estimates to $\delta \log(M_\star) \approx 0.3$ dex for galaxies with $M_\star > 10^9 M_\odot$ at $z \approx 0.5$, and offers improvement compared to optical-only estimates out to $z \approx 1$, with stellar masses constrained within a factor of approximately 2.5. When combined with other multi-wavelength imaging of the Stripe, including moderate-to-deep ultraviolet (*GALEX*), optical and mid-infrared (*Spitzer*-IRAC) coverage, as well as tens of thousands of spectroscopic redshifts, VICS82 gives access to approximately 0.5 Gpc$^3$ of comoving volume. Some of the main science drivers of VICS82 include (a) measuring the stellar mass function of $L^\star$ galaxies out to $z \sim 1$; (b) detecting intermediate redshift quasars at $2 \lesssim z \lesssim 3.5$; (c) measuring the stellar mass function and baryon census of clusters of galaxies, and (d) performing optical/near-infrared–cosmic microwave background lensing cross-correlation experiments linking stellar mass to large-scale dark matter structure. Here we define and describe the survey, highlight some early science results and present the first public data release, which includes an SDSS-matched catalogue as well as the calibrated pixel data itself.

*Keywords:* surveys – catalogs – infrared: general

## 1. INTRODUCING VICS82

Extragalactic, and indeed Galactic, astronomy has entered an era of deep large-area surveys. This has been facilitated by improvements in instrumentation such as large format cameras that can efficiently map huge swathes of sky with great sensitivity, coupled with the use of dedicated survey telescopes. This theme will shape the research landscape during the coming decades, with several giant surveys coming online now and in the near-future that will survey significant fractions of the sky in the optical and near-infrared (e.g., Pan-STARRS[13], Dark Energy Survey[14], Hyper-SuprimeCam[15], J-PAS[16], Large Synoptic Survey Telescope[17], *Euclid*[18]) and in the radio bands (e.g., Square Kilometer Array pathfinders, LOFAR[19]). On the other hand, there will remain a need for ultra-deep "keyhole" multi-wavelength surveys that can hunt galaxies in the very early Universe.

The familiar "deep extragalactic survey fields", such as the Great Observatories Origins Deep Surveys (GOODS; Giavalisco et al. 2004), established through significant observational investment over more than a decade have been the key resources from which much of our understanding of the high redshift Universe has been gleaned. These deep surveys typically cover areas of no more than a square degree and their pencil-beam nature naturally trade off volume for depth. A clear niche is the intermediate-scale (of order 100 square degree) survey that balances the statistical benefits of large area coverage with moderately deep multi-wavelength coverage. With its unique combination of imaging and spectroscopic components, over the past decade the Sloan Digital Sky Survey (SDSS, York et al. 2000) has revolutionised studies of galaxy populations and large scale structure in the local ($z < 0.3$) Universe. The ability to perform studies with similar statistical accuracy at higher-$z$ would represent a dramatic step forward in our understanding of the evolution of the galaxy populations of the early Universe.

During the fall seasons of 2000–2007, the SDSS repeat-edly scanned a narrow (2.5 degrees in declination) 270 square degree strip along the celestial equator. In SDSS nomenclature this region is known as 'Stripe 82'. The optical imaging depth ($i \simeq 22.8$, $z \simeq 21.8$ AB) is two magnitudes fainter than the main SDSS survey (Annis et al. 2014), providing a deep probe of the Galactic structure, the evolution of galaxy populations to $z \approx 1$, and the demographics of faint and distant quasars.

[1] Center for Astrophysics Research, Science & Technology Research Institute, University of Hertfordshire, Hatfield, AL10 9AB, UK. j.geach@herts.ac.uk
[2] Institute of Astronomy and Astrophysics, Academia Sinica, Taipei 10617, Taiwan
[3] Centro Brasileiro de Pesquisas Físicas, Rua Dr. Xavier Sigaud 150, CEP 22290-180, Rio de Janeiro, RJ, Brazil
[4] Laboratoire d'Astrophysique, Ecole Polytechnique Fédérale de Lausanne, Observatoire de Sauverny, CH-1290 Versoix, Switzerland
[5] Laboratoire d'Astrophysique de Marseille, UMR 7326, F-13388 Marseille, France
[6] Institute for Astronomy, University of Edinburgh, Royal Observatory, Edinburgh, EH9 3HJ, UK
[7] University of California Observatories, UC Santa Cruz, 1156 High St, Santa Cruz, CA, 95064
[8] Institut d'Astrophysique de Paris, 98bis Boulevard Arago, F-75014 PARIS, France
[9] Departamento de física teórica, universidad autónoma de Madrid
[10] Dipartimento di Fisica e Scienze della Terra, Università degli Studi di Ferrara, Via Saragat 1, I-44122 Ferrara, Italy
[11] Department of Physics and Astronomy, University of British Columbia, 6224 Agricultural Road, Vancouver, BC V6T 1Z1, Canada
[12] Observatorio Astronómico de Córdoba, Laprida 854, Córdoba Capital, Argentina

[13] http://pan-starrs.ifa.hawaii.edu/public/
[14] https://www.darkenergysurvey.org/
[15] http://hsc.mtk.nao.ac.jp/ssp/
[16] http://www.j-pas.org/
[17] https://www.lsst.org/
[18] http://sci.esa.int/euclid/
[19] http://www.lofar.org/

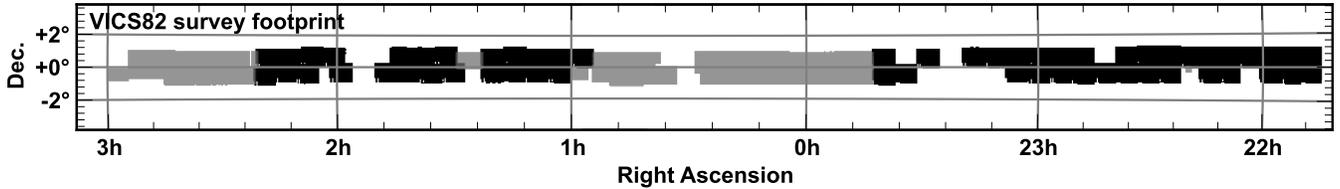

**Figure 1.** VICS82 survey footprint. Grey regions indicate CFHT coverage, black regions indicate VISTA coverage. Gaps in the coverage indicate where a pointing was omitted due to the presences of a very bright (7th magnitude or brighter) star, or where the survey is incomplete.

The wealth of multi-wavelength data in Stripe 82 is unparalleled among extragalactic fields of comparable size. Stripe 82 already has a high density of spectroscopy, with tens of thousands of redshift measurements from SDSS, 2SLAQ (Richards et al. 2005), 2dF (Colless et al. 2001), 6dF (Jones et al. 2004), DEEP2 (Newman et al. 2013), VVDS (Le Fèvre et al. 2005), and PRIMUS (Coil et al. 2011). Surveys such as the SDSS-III Baryon Oscillation Spectroscopic Survey (BOSS; Dawson et al. 2013), SDSS-IV/eBOSS(SDSS Collaboration et al. 2016), WiggleZ (Drinkwater et al. 2010) and soon the Hobby-Eberly Telescope Dark Energy Experiment[20] (HETDEX) have added and will add tens of thousands more spectra to this legacy. In addition to imaging in the Sloan bands, the Stripe is covered by *GALEX* Far/Near UV imaging, with exposure twice the depth of the *GALEX* Medium Imaging Survey (Morrissey et al. 2007), and by the United Kingdom Infrared Deep Sky Survey (UKIDSS; Lawrence et al. 2007) Large Area Survey (LAS) in the YJHK bands (to $K = 20.2$) which are photometrically matched to the SDSS coadd photometry in Bundy et al. (2015). Recently, an area of 160 deg² of the Stripe has been imaged by the Canada-France-Hawaii Telescope (CFHT) Stripe 82 Survey (CS82) in the $i'$-band down to $i' \simeq 24.1$ with a median seeing FWHM of $0.6''$ (Erben et al. 2013), allowing for precision weak lensing measurements (Liu et al. 2015; Battaglia et al. 2016). Stripe 82 has also been observed as part of DES and the S-PLUS[21] (Oliveira et al. in preparation). A 31.3 deg² area of the Stripe has been covered by *Chandra* and, mostly, *XMM-Newton*, for the Stripe 82 X-ray survey (LaMassa et al. 2016).

Degree-scale sub-regions also overlap with deeper imaging: the UKIDSS Deep eXtragalctic Survey (DXS) and CFHT Legacy Survey (CFHTLS) W4 fields. In the mid-infrared the *Spitzer* HETDEX Exploratory Large-area Survey (SHELA, Papovich et al. 2016) and the *Spitzer* IRAC Equatorial Survey (SpIES, Timlin et al. 2016) have obtained $3.6\mu m$ and $4.5\mu m$ imaging of 24 deg² and 115 deg² regions of Stripe 82 respectively to $5\sigma$ depths of $\sim 5\,\mu$Jy.

At longer wavelengths, there have been two major *Herschel* surveys covering the Stripe, with the *Herschel* Stripe 82 Survey (HerS) obtaining 250, 350 and 500$\mu m$ imaging of 79 deg² of the Stripe to depths of 13.0, 12.9, and 14.8 mJy beam$^{-1}$ (Viero et al. 2014) and the HerMES Large Mode Sur-vey (HeLMS) has covered 274 deg² of SPIRE imaging, also overlapping with the Stripe (Asboth et al. 2016). The full Stripe also lies within the footprint of Atacama Cosmology Telescope (ACT; Fowler et al. 2010) equatorial survey (r.m.s. 23 $\mu$K-arcmin at 148 GHz), and 80 square degrees of the Stripe has 1.4 GHz Very Large Array (VLA) imaging three times deeper than the VLA FIRST survey that reaches a typical r.m.s. depth of 0.15 mJy at 1.4 GHz (Hodge et al. 2011).

Stripe 82 is emerging as the first of a new generation of $\Omega > 100$ deg² deep extragalactic survey fields, with an impressive array of multi-wavelength observations already in-hand or in progress. Here we present VICS82: the VISTA–CFHT Stripe 82 survey covering approximately 150 square degrees of the Stripe to a depth of $J \approx 22$ mag and $K_s \approx 21.5$ mag (AB), a valuable addition to the growing legacy of data in this field. In this article we describe the survey and present the first data release. In §2 we describe the field layout, observations, details on calibration and data reduction strategy, and source extraction (including key dianostics such as image quality and depth). In §3 we outline our main science goals and summarise the survey in §4. Throughout we give magnitudes on the AB system (Oke & Gunn 1983) unless otherwise stated.

**Table 1**
Central wavelengths and bandpasses of near-infrared filters used in VICS82.

| Filter | Central wavelength ($\mu$m) | Bandpass ($\mu$m) |
|---|---|---|
| | CFHT–WIRCam | |
| $J$ | 1.25 | 0.16 |
| $K_s$ | 2.15 | 0.33 |
| | VISTA–VIRCAM | |
| $J$ | 1.25 | 0.18 |
| $K_s$ | 2.15 | 0.30 |

## 2. THE VICS82 SURVEY

VICS82 is conducted with the Canada France Hawaii Telescope (CFHT) WIRCam instrument and with the Visible Infrared Survey Telescope for Astronomy (VISTA) VIRCAM instrument; survey load was split between the facilities. VICS82 is a $J$ and $K_s$-band survey, with the VISTA and CFHT broadband filters well-matched (Jarvis et al. 2013; Table 1). Here we describe the observation strategy, data reduction, calibration and source extraction methods.

### 2.1. Field layout and observation strategy

The VICS82 coverage of Stripe 82 is a near-contiguous $\sim 150$ deg² region defined by the boundaries 3 hours and 22.2 hours in Right Ascension and $-1 < \delta < 1$ degrees in Declination (Figure 1). With the combination of VISTA/VIRCAM and CFHT/WIRCam pointings we obtain nearly uniform coverage, however there are gaps in the tiling strategy where we have avoided several bright (<7th magnitude) stars. The target depth was 22 mag in both $J$ and $K_s$ bands across the survey footprint. Each WIRCam 'tile' is a $3 \times 3$ mosaic of individual $21.5' \times 21.5'$ WIRCam pointings, and the VISTA 'tiles' are single VIRCAM $1° \times 1°$ pointings. In total, VICS82 is constructed from 33 VIRCAM tiles and 55 WIRCam tiles.

---
[20] http://hetdex.org/
[21] http://www.iag.usp.br/labcosmos/en/s-plus/

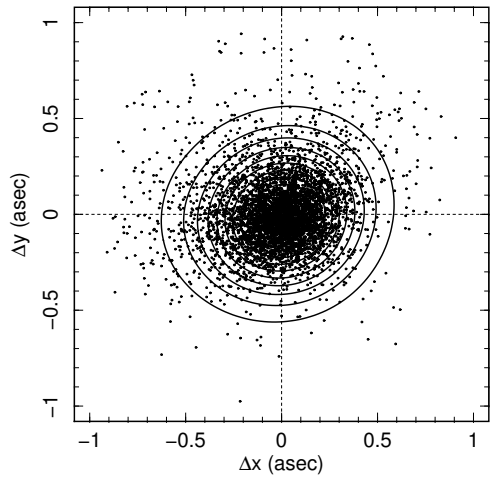

**Figure 2.** Accuracy of the absolute astrometric calibration of VICS82, relative to 2MASS for unsaturated point sources matched within $1''$. Points show the relative offset in Right Ascension and Declination of point sources extracted from a full VICS82 tile, compared to their counterpart in 2MASS. Contours simply visualise the density of points in a smoothed kernel representation. The mean offset is consistent with zero, with standard deviation in the distribution of offsets in each direction: $\Delta\alpha \approx 0.18''$, $\Delta\delta \approx 0.15''$. Note that the internal positional accuracy is much better than this, with a residual of $0.1''$ at the $5\sigma$ limit of the survey and less than $0.05''$ at $K_s < 20$ mag (see §2.3).

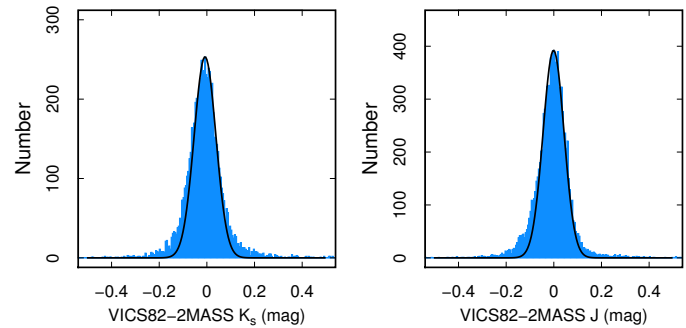

**Figure 3.** Comparison of 2MASS and VICS82 $J$ and $K_s$ magnitudes for sources with total magnitudes in the range $K_s = 15$–$15.25$ mag. Each histogram contains a total of $10^4$ randomly selected sources across the survey. The width ($\sigma$) of each distribution is 0.044 and 0.046 mag respectively, where we have fit a gaussian to the central distributions after clipping. Note that these numbers are averaged over *both* CFHT and VISTA photometry across the *full* VICS82 survey area.

To obtain the required VISTA/VIRCAM integration times of 180 s and 200 s in $J$ and $K_s$ respectively, observations used Detector Integration Times (DITs) of 10 s per exposure with on-chip NDITs of 9 and 10 in $J$ and $K_s$. The standard 6-point dither pattern fills gaps between detectors, producing effectively an image of $1.45 \times 1.05$ deg$^2$ per observation when covering wide areas. For the CFHT observations the standard WIRCam 9-point dither pattern is used to cover chip gaps and average over bad pixels. Individual exposure times are 55 s and 20 s in $J$ and $K_s$ to build up frame integrations of 330 s and 180 s respectively over fields of $21' \times 21'$. To evaluate the total non-overlapping area of the survey we create a block-averaged mosaic of the VIRCAM And WIRCam tiles and sum the number of pixels then multiply by the pixel area. This yields approximately 150 square degrees.

### 2.2. *Data reduction and calibration*
#### 2.2.1. *VISTA VIRCAM*

The raw images are pre-processed (detrending involving dark and flat field correction, first pass sky subtraction, astrometric and photometric calibration) by the Cambridge Astronomical Survey Unit (CASU), and subsequent processing (refined sky subtraction, astrometric solutions, stacking and quality control) are performed at TERAPIX. The pre-reduction steps are identical to that of the UltraVISTA deep Survey (McCracken et al. 2012). All calibration frames (sky, flat and bad pixel masks) are processed by CASU using the VIRCAM version 1.3 release. Science images were obtained between October 2nd 2012 and January 14th 2013. Each science images is graded based on the ESOGRADE keyword; we rejected all images with grade C (158 images). The individual pre-processed images from CASU were used to identify and flag the saturated pixels, and these maps are used to discard saturated objects from subsequent catalogs because they degrade the accuracy of the SCAMP (Bertin 2006) astrometric and photometric calibrations (see below).

The QualityFITS software is applied to all input data, producing weight-maps and catalogs and providing an initial quality assessment beyond the rejection of grade C images described above. We compute astrometric and photometric solutions with SCAMP, using LDAC (Leiden Data Analysis Center) catalogs produced by QualityFITS. The reference catalog for the astrometric and photometric absolute calibration is 2MASS (Skrutskie et al. 2006). SWarp (Bertin 2010) is used to combine the individual pre-reduction images and weight maps using the astrometric solution from SCAMP. This stack is then used to produce a binary mask for the 'proper' sky-subtraction step. The use of a deep stack to create the mask instead of single exposure images enables a complete removal of all faint objects, including those not detected in a single exposure.

To create sky-subtracted images we start by adding to each image the sky background frames originally subtracted by CASU; this recovers the detrended images without sky subtraction. Based on the first pass stack and astrometric solutions, we then compute object masks for each individual image. We use these object masks (appropriately warped to match the images) to compute and subtract a running sky for each individual image based on a median of images taken during a 20 minute interval. After the subtraction of the running sky, we 'destripe' the images in both directions and remove large-scale background gradients using SExtractor (Bertin & Arnouts 1996). After this step the images are visually inspected to isolate problems that could persist after the sky subtraction process. Images with poor sky subtraction and/or unacceptable residuals (cosmetic defects, large scale patterns, etc.) are eventually rejected through visual inspection. Finally, SCAMP is used to compute the astrometric and relative photometric calibration (field to field rescaling) using the LDAC catalogs produced in the earlier step by QualityFITS.

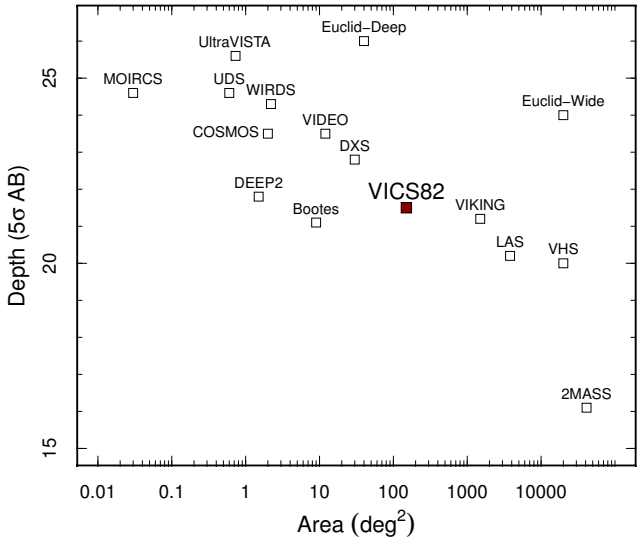

**Figure 4.** A comparison of the limiting depths and areas of various near-infrared surveys, both existing and planned. Note that the *Euclid* mission will not observe broadband $K$: imaging photometry will be in the $Y$, $J$ and $H$ bands, reaching a depth of 24 mag in each for the wide survey, and two magnitudes deeper in a smaller (40 deg$^2$) deep survey. VICS82 fills a niche in the area–depth parameter space linking small, deep surveys with those that cover much larger solid angles to shallower depth.

Photometric calibration is checked by comparing the magnitudes measured on the images with the corresponding photometry in the 2MASS catalog.

### 2.2.2. CFHT WIRCam

The WIRCam data are reduced with the SIMPLE Imaging and Mosaicking PipeLinE (SIMPLE, Wang 2010) under the Interactive Data Language (IDL) environment. The WIRCam raw images are first corrected for nonlinearity, and then images that were taken in the same dither sequence and from the same HAWAII2-RG chips are then grouped and reduced together. Images are self-flattened with a two-pass procedure: grouped images are first normalized and median-combined to form a sky flat. Objects detected in the flattened images are masked from the original images, and these masked images are again normalized and median-combined to form a cleaner final sky flat. On the flattened and object-masked images a background is fitted with a 5th-degree polynomial surface, which is subtracted from the image to improve flatness.

The flattened and sky-subtracted images are then corrected for distortion and astrometry. An initial distortion correction is derived from the changes of positions of detected objects in the dithered images (see Wang 2010 for details). The final astrometry calibration and projection of the images, which also include the correction of distortion, are made by matching the positions of detected objects to their coordinates in the 2MASS point-source catalog. After astrometric calibration the images are coadded; these images are flux calibrated by comparing the source fluxes (measured with 5″ diameter apertures) with the 2MASS point-source catalog. Only objects with Vega magnitudes in the ranges of $J = 14$–$16$ and $K_S = 12.6$–$14.5$ are used for flux calibration to avoid effects of nonlinearity in the WIRCam images (bright end) and selection effects in the 2MASS catalog (faint end). We adopt the 2MASS 'default magnitudes' which attempt to account for the total fluxes of the point sources. Finally, coadded and flux-calibrated images from different chips and from different dither sets are further combined to form deep wide-field mosaic images (tiles).

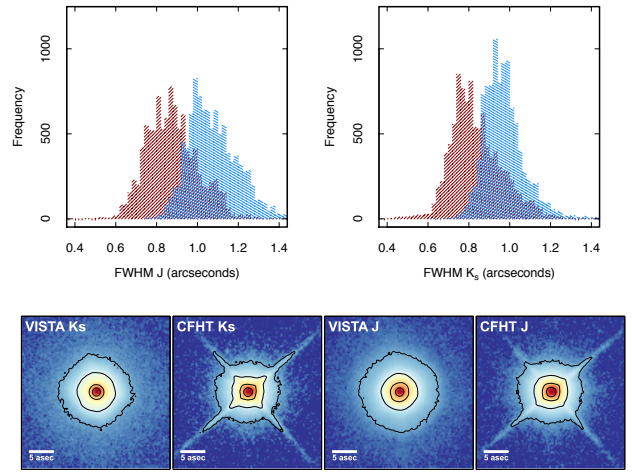

**Figure 5.** Distribution of average image quality for VICS82, evaluated as the full width at half maximum of $10^4$ point sources extracted from randomly selected image tiles for each of the CFHT (blue) and VISTA (red) observations. We achieve sub-arcsecond seeing across the majority survey in both $J$ and $K_s$ bands (Table 2), with systematically better image quality in the VISTA tiles. The lower panels show the median average images of the $10^4$ sources for each telescope and band. Each panel is 45″ on a side (the scale bar shows 5″), and the images are normalized to unity at peak. The contours are at levels of 0.01, 0.1, 1 and 10% of peak. These average PSFs are used in the completeness simulation described in §2.6.

### 2.3. Calibration

Figure 2 illustrates the typical accuracy of the VICS82 absolute astrometric calibration relative to 2MASS. The mean relative positional offset is consistent with zero, with a spread (standard deviation) of less than 0.2″ in Right Ascension and Declination; this can be taken as the typical uncertainty in the absolute astrometric calibration. The relative astrometry is far more accurate, and we measure this using the completeness simulation described in §2.6: by injecting point sources at a known position and then measuring the offset from recovered (detected) position, we can assess the typical astrometric uncertainty as the root mean squared offset as a function of flux. This injection-recovery process is explained in more detail in §2.6 where we use it to assess survey depth and completeness. At $K_s = 21.5$ mag we measure an r.m.s offset of 0.1″ between input and recovered position (symmetric in Right Ascension and Declination), falling to less than 0.05″ for $K_s < 20$ mag. We confirm that this level of accuracy is consistent across the full survey region. Figure 3 shows the dispersion in the photometric calibration for the $J$ and $K_s$-bands respectively for unsaturated sources with total magnitudes of 15–15.25 mag. Again, 2MASS–VICS82 photometry residuals are consistent with zero, and the dispersion is 0.044 mag in $J$ and 0.046 mag in $K_s$; this result is also consistent across the full survey. Note that the VISTA–2MASS calibration contains a color term (McCracken et al. 2012), but no color term is available for the WIRCam–2MASS calibration.

### 2.4. Source extraction

**Table 2**
Average image quality of VICS82. Average seeing is defined by the median FWHM of gaussian fits to 10,000 point sources across the entire survey area in each filter and from each telescope. The uncertainty is the standard deviation of the FWHM distributions (Figure 5).

| Telescope | $K_s$ | $J$ |
|---|---|---|
| CFHT | $(0.96 \pm 0.10)''$ | $(1.06 \pm 0.12)''$ |
| VISTA | $(0.82 \pm 0.13)''$ | $(0.87 \pm 0.13)''$ |

We use SExtractor (version 2.14.7) to perform source detection, extraction and photometry. For both CFHT and VISTA, we extract using a weight image derived during the data reduction with a 'vanilla' parameter set, the main components of which are a 'detection threshold' (DETECT_THRESH) of 2 with a 'minimum area' (MIN_AREA) of 3 contiguous pixels meeting the detection threshold. Inspection of the source catalogues reveals that this set of parameters is effective at detecting the widest range of sources (from the low signal-to-noise regime to the brightest extended sources) with sensible de-blending of unassociated emission and low contamination from obviously spurious sources. The final catalogue, after rejecting duplicate detections across overlapping tiles, contains 9.5 million sources with $K_s < 22$ mag across a total of 150 square degrees. In Figure 8 we plot the $J$ and $K_s$-band galaxy number counts, corrected for completeness (§2.6), compared to other surveys. There is excellent agreement with data from the literature down to the survey depth (§2.6). Note that we have rejected stars using the same $(g-i)$–$(J-K_s)$ stellar locus definition as Jarvis et al. (2013), which is based on Baldry et al. (2010).

### 2.5. *Image quality*

We assess image quality across the survey by measuring the FWHM of bright unsaturated point sources with $14 < K_s < 15$ mag detected in the catalogue described above, retaining sources with CLASS_STAR (a measure of star/galaxy separation) > 0.95. In addition, we create a median stack of the point sources (normalising each source to its peak flux) to generate an average PSF for each band and telescope. In Figure 5 we show the distribution of image quality for the full survey, and the average PSFs derived from the stacking. We use the latter to derive aperture corrections for photometry through a simple curve-of-growth analysis. Table 2 summarises the average values for the image quality, which is clearly systematically better in the VISTA imaging, but note that we achieve $\lesssim 1''$ seeing across the majority of the survey in both bands and telescopes. Table 3 lists the aperture-to-total flux corrections for apertures with diameters 1–5″ derived from a curve-of-growth analysis of the average PSFs shown in Figure 5.

### 2.6. *Survey depth and completeness*

To evaluate survey depth and completeness, we run a simple simulation where model point sources of varying total magnitude are inserted into the images and then re-extracted; the rate of recovery of these sources allows us to estimate survey completeness and a measure of depth. We use the average stack PSFs described above (§2.5, Figure 5) scaled such that their magnitude measured in a 2″ diameter aperture is $20 < K_s < 24$ mag. At each flux interval 10,000 sources are inserted into the data, with each model source added at a random point within a randomly chosen tile from the survey. SExtractor is then used to recover these sources, adopting the same detection criteria as used in our 'real' catalogue extraction described above. Figure 6 shows the completeness curves for the $J$ and $K_s$ bands for each telescope. We also determine the average signal-to-noise ratio of the flux measured in a 2″ aperture for each recovered source, with the $5\sigma$ limit approximately corresponding to the 80% completeness level.

**Table 3**
Aperture corrections to total magnitudes for point sources in VICS82.

| Diameter (arcseconds) | CFHT (mag) | | VISTA (mag) | |
|---|---|---|---|---|
| | $J$ | $K_s$ | $J$ | $K_s$ |
| 1.0 | −0.94 | −0.81 | −0.84 | −0.71 |
| 1.5 | −0.43 | −0.38 | −0.46 | −0.36 |
| 2.0 | −0.26 | −0.24 | −0.33 | −0.25 |
| 2.5 | −0.18 | −0.18 | −0.26 | −0.20 |
| 3.0 | −0.13 | −0.14 | −0.21 | −0.16 |
| 3.5 | −0.10 | −0.11 | −0.17 | −0.14 |
| 4.0 | −0.08 | −0.09 | −0.14 | −0.13 |
| 4.5 | −0.07 | −0.08 | −0.11 | −0.11 |
| 5.0 | −0.06 | −0.07 | −0.09 | −0.10 |

**Table 4**
Point source depth and completeness limits for VICS82. We report the average $5\sigma$ detection threshold (measured in a 2″ aperture, corrected to total) and 50% and 80% completeness limits for the CFHT and VISTA images following the methods described in §2.3. The parameters $m_{50}$ and $\mathcal{F}$ can be used in equation 1 to model the completeness as a function of magnitude.

| Telescope | $m_{50}$ (mag) | $m_{80}$ (mag) | $\mathcal{F}$ | $5\sigma$ depth (mag) |
|---|---|---|---|---|
| | | $K_s$ | | |
| CFHT | 21.9 | 21.4 | 3.11 | 20.9 |
| VISTA | 21.8 | 21.4 | 3.24 | 20.9 |
| | | $J$ | | |
| CFHT | 22.4 | 21.9 | 2.98 | 21.4 |
| VISTA | 22.4 | 21.9 | 2.98 | 21.5 |

Each completeness curve can be modelled by a smoothed step function of the form

$$C = \frac{1}{1 + e^{\mathcal{F}(m - m_{50})}} \quad (1)$$

where $C$ is the completeness, $m_{50}$ is the 50% completeness level and $\mathcal{F}$ is the smoothing parameter. This function provides a convenient analytic form to model survey completeness as a function of magnitude. The average 50% and 80% completeness and $5\sigma$ point source depths (aperture corrected) are given in Table 4, along with the parameter $\mathcal{F}$.

### 2.7. *Reliability*

We evaluate the false detection rate as a function of magnitude by running SExtractor (as in §2.4) on $10^5$ randomly selected $5' \times 5'$ regions of each of the CFHT and VISTA parts of the survey. First, the source extraction is run on the image to determine the locations of positive sources; the corresponding pixels are then set to the background median of the image. This image is then inverted and the source extraction is run again with the same parameters; for gaussian noise, the number of detected 'sources' in the inverted image will correspond to the expected number of false positives in the real catalogue.

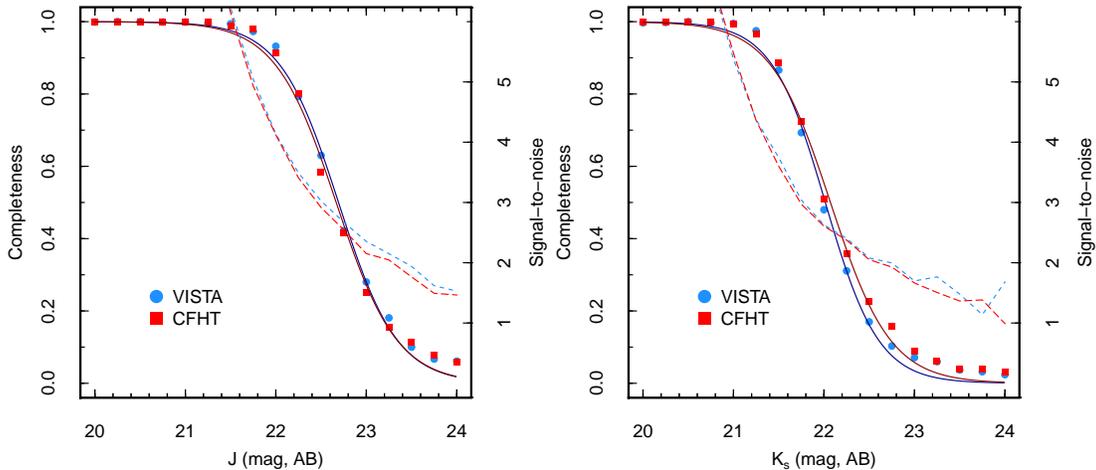

**Figure 6.** Completeness curves for VICS82, split by band and telescope. Completeness rates are determined by injecting model point sources into the data, scaled by 2″ aperture flux, and then attempting to recover using the same source detection criteria as the real catalogue (§2.6). The dashed lines show the median signal-to-noise ratio of the 2″ aperture flux for injected sources, demonstrating that the 80% completeness level corresponds approximately to a 5$\sigma$ detection; we take this as the formal survey point source limit. Table 4 lists the completeness limits and 5$\sigma$ magnitudes for each telescope and band.

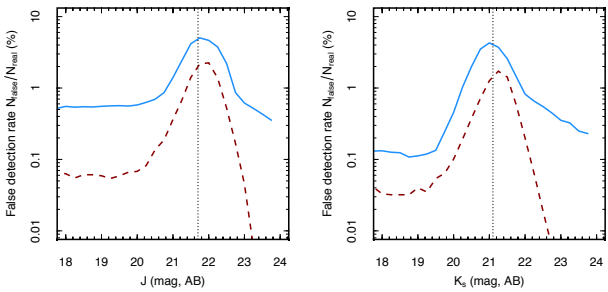

**Figure 7.** False detection rates $N_{\rm false}/N_{\rm real}$ as a function of magnitude. The false detection rates were determined by inverting sub-images and running the source extraction as described in §2.4 (further details of the method in §2.7). Any 'sources' detected in these inverted images are noise spikes meeting the detection criteria; for gaussian noise this will correspond to the expected number of false positives at a given flux level in the real catalogue. The vertical line in each panel marks the average 5$\sigma$ limit of the survey in each band. The false detection rates at this limit are 2.3% (CFHT, dashed line) and 4.6% (VISTA, solid line) in $J$ and 1.4% (CFHT) and 2.6% (VISTA) in $K_{\rm s}$.

Figure 7 shows the false detection rate as a function of magnitude, illustrating a characteristic rise in false positives as one approaches the survey limit. At the average 5$\sigma$ point source limit the false detection rate in the $J$ band is 4.6% and 2.3% for the VISTA and CFHT catalogues respectively. In the $K_{\rm s}$ band the rates are 2.6% and 1.4% at the survey limit. Note that these figures no not reflect the increased false detection rate due to spurious sources resulting from (for example) bright halos and diffraction spikes around stars.

### 2.8. *First data release*

Our intention is to deliver a series of data releases of increasing sophistication, culminating in a fully band-merged catalogue with optimally homogenised (PSF-matched) optical–infrared photometry and 'added value' data products including $ugrizJK_s$+3.6$\mu$m+4.5$\mu$m photometric redshifts incorporating the VICS82 photometry with existing SDSS optical data and mid-infrared photometry from *Spitzer*/IRAC (Papovich et al. 2016; Timlin et al. 2016). In this first VICS82 data release (DR1) we provide a $K_s$-selected catalogue, cut at $K_s$ = 22 mag, matched to the independent $J$ catalogue. This catalogue contains about 9.5 million sources over 151 square degrees, with 'total' (MAG_AUTO) and aperture (1″, 1.5″, 2″, 2.5″ and 3″ diameter, MAG_APER) magnitudes derived from our 'vanilla' extraction procedure described above (§2.4). The catalogue has been purged of internal matches (for overlapping tiles) using a 1″ elimination radius (roughly 3$\sigma$ in terms of the astrometric uncertainty). We also match the VICS82 catalogue to the SDSS DR9 catalogue using a 2″ matching radius, providing SDSS *ugriz* (deep coadd) photometry, spectroscopic redshifts and classifications. In addition to the catalogue, we provide access to the calibrated pixel data via a cut-out server available at **http://stri-cluster.herts.ac.uk/vics82**. Full image tiles are also available from this URL.

## 3. OVERVIEW OF SELECT SCIENCE GOALS

VICS82 stands in a unique position in the depth-area parameter space of existing near-infrared surveys (Figure 3). There is diverse science potential for this data; as described above, we will make timely data releases of calibrated imaging and catalogue products of increasing sophistication that can be used by the community, but we have some specific science goals (that originally motivated the survey) that we briefly overview here.

### 3.1. *Stellar mass functions of $L > L_\star$ galaxies to $z \sim 1$*

Sampling the rest-frame $J$-band at $z \sim 1$, $K_s$-band imaging has the potential to improve photometric redshift estimates and stellar mass estimates over what can be achieved when the reddest available filter is the $z$-band, which is blueward of the 4000Å break at $z > 1$. To examine the quality of photometric redshift estimates based on the Stripe 82 coadd optical photometry after adding the VICS82 photometry, we derive photometric redshifts ($z_{\rm phot}$) for VICS82 objects with existing spectroscopic redshifts. Photometric redshifts are measured using the *DEmP* code (Hsieh & Yee 2014).

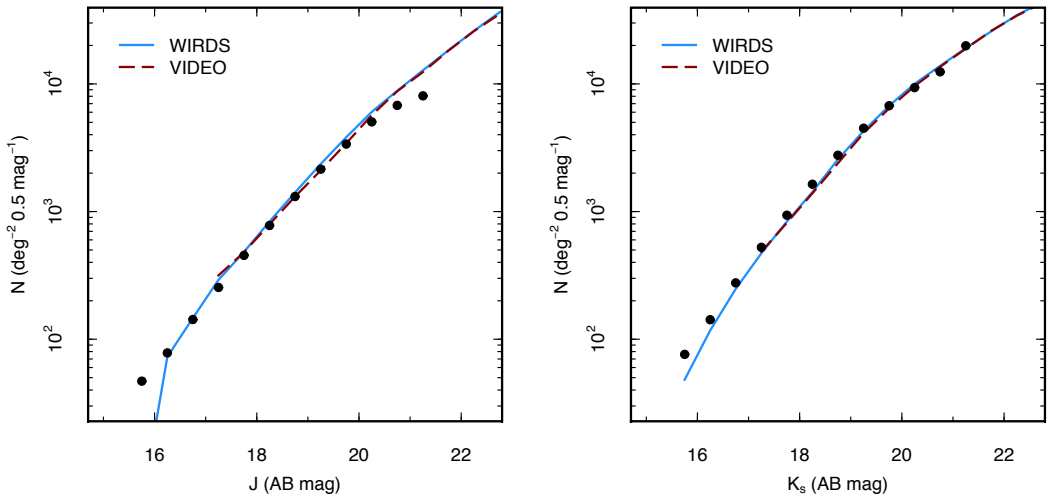

**Figure 8.** Number counts of galaxies detected in VICS82 in $J$ and $K_s$, not corrected for incompleteness. We compare the counts to those measured in the same bands in the deeper near-infrared surveys VIDEO (Jarvis et al. 2013) and WIRDS (Bielby et al. 2012), with excellent agreement to the VICS82 survey depth. Note that the slight incompleteness in the $J$-band counts at the faint end is due to the construction of the master catalogue, which is effectively $K_s$-selected.

Although the filter transmission curves of VISTA/VIRCAM and CFHT/WIRCam are very similar, the slight difference can still affect the quality of the photometric redshift if the two filter systems are assumed to be identical in the photometric redshift code. Therefore, the photometric redshifts of the VISTA/VIRCAM and CFHT/WIRCam photometry are derived separately. We first match the VICS82 catalogue to various available spectroscopic redshift catalogues (mainly BOSS). The matched catalogues include 40112 objects (in the VISTA footprint) and 32,706 objects (in the CFHT footprint). Because *DEmP* is an empirical photometric redshift code, a training set is needed. We therefore split each matched catalogue in half, and use one as the training set, the other the validation set. We first perform the photometric redshift estimation for the validation set using the SDSS photometry only, then repeat the same procedure using the SDSS + VICS82 photometry. The results are shown in Figure 9 (left panel), where we compare $(z_{\rm phot}-z_{\rm spec})/(1+z_{\rm spec})$ versus true redshift for the SDSS-only and SDSS+VICS82 fits, with an error bar that shows the $1\sigma$ standard deviation of the redshift residual to represent the scatter.

Both fits give residuals consistent with zero, with a scatter that increases significantly beyond $z > 0.8$. Interestingly, the addition of the VICS82 does not significantly improve the accuracy of photometric redshifts compared to the optical photometry alone. A possible reason is that the main features driving the photometric redshift fit (e.g. the 4000Å break) are still in the optical bands at $z < 1$. At $z > 1$, where we might expect gains in the photometric redshift fitting when including VICS82 photometry, the SDSS depth starts to become important in the signal-to-noise ratio of high-$z$ sources. In this case, the VICS82 data should provide a greater improvement on photometric redshifts when matched to DES or the forthcoming J-PAS data on Stripe 82.

In addition to the photometric redshift, we also examine the quality of the stellar mass estimate when the VICS82 photometry is included. Doing so requires that the "true" stellar masses for objects in the test sample is known. To proceed, we thus assume that the stellar mass of an object derived using SED (spectral energy distribution) fitting with the SDSS and VICS82 photometry and its spectroscopic redshift is the true answer. We use *newhyperz* version 11[22] with the GALAXEV stellar synthesis model (Bruzual & Charlot 2003) to perform the SED fitting. To test how the accuracy of the stellar mass estimate can be improved by adding the VICS82 photometry, we perform the SED fitting for the validation set twice. The first run uses the SDSS photometry only with the photometric redshift derived using the SDSS photometry. The second run uses the SDSS + VICS82 photometry with the photometric redshift derived using the SDSS + VICS82 photometry. The results are shown in Figure 9 (right panel). Here we see significant improvements in the the stellar mass estimate with VICS82 photometry is used: the scatter in $\log_{10}(M_{\rm photo-z}/M_{\rm reference})$ reduces by a factor $\sim 2$ at $z \approx 0.5$ to $\sim 0.3$ dex. Note also that SDSS-only mass estimates are systematically biased high by $\sim 0.1$–$0.2$ dex. At higher redshifts the scatter starts to increase in the stellar mass estimate, but with the SDSS+VICS82 fits systematically improved over SDSS alone. On average, the scatter in the residual for stellar mass estimates at $z < 1$ reduces from $\sim 0.7$ dex to $\sim 0.4$ dex for galaxies with $M_\star > 10^9 M_\odot$ when VICS82 photometry is added to the SDSS optical photometry.

Large survey fields usually lack comprehensive spectroscopic follow-up: VICS82 offers the distinct advantage that the majority of galaxies with stellar mass $M_\star > 10^{11} M_\odot$ have spectroscopic redshifts from BOSS (also forming an excellent training set for precision photometric redshifts of lower-mass galaxies). The fainter optical magnitudes of galaxies accessible at the VICS82 depth will be increasingly well measured by DES and HSC surveys in the field. The 1–2% statistical precision in number densities measured in the 0.5 Gpc$^3$ VICS82 volume ($0.3 < z < 1.2$) enables us to address some key issues: (a) the basic prediction of hierarchical mass assembly in the $\Lambda$CDM framework that has yet to be verified in measurements of the evolving abundance of massive galaxies; (b) tracking the flow of evolving populations by measuring how the declining number density of one category is compensated by the rise of another; (c) using morphological data (e.g. bulge-to-disc ratios, half light radii, Sérsic indices and more sophisticated surface brightness fitting algorithms) from the existing CFHT *i*-band data (Moraes et al, in preparation) and star formation rates. VICS82 can also explore the processes that drive star formation quenching and the formation of bulge-dominated galaxies, linking such populations with their progenitors (Bundy et al. 2010).

---
[22] http://userpages.irap.omp.eu/~rpello/newhyperz/

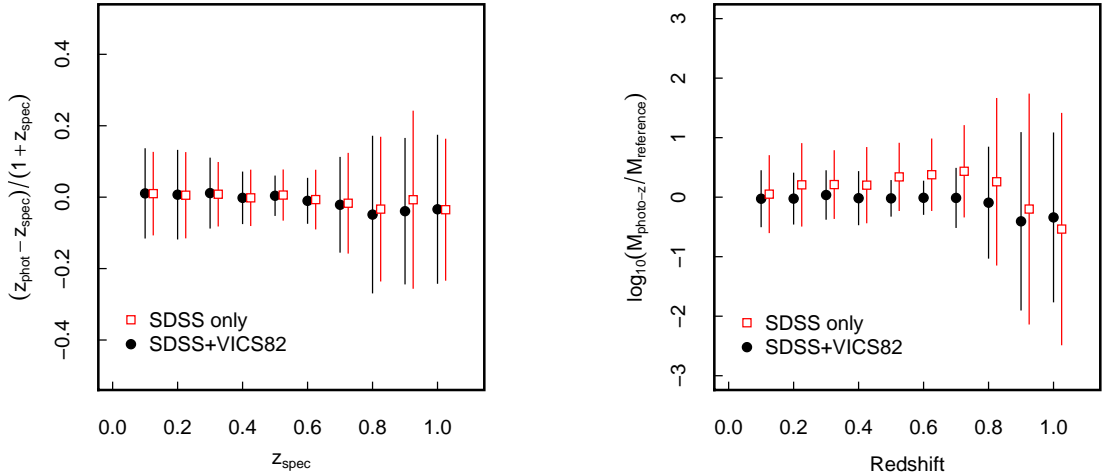

**Figure 9.** (left) Photometric redshift accuracy for galaxies as a function of redshift, comparing fits using just SDSS (Stripe 82 coadd) photometry and the combination of SDSS+VICS82. The error bars reflect the 1$\sigma$ standard deviation in the residual. Since key photometric fitting features such as the 4000Å break are still in the optical bands out to $z \approx 1$, it is likely the reason (in combination with the SDSS Stripe 82 optical depth) the scatter remains similar. (right) The improvement in stellar mass estimates (using photometric redshifts) when VICS82 photometry is added to SDSS, where the reference mass is derived from the best fitting SED template for galaxies where the spectroscopic redshift is known. We make a cut in stellar mass $>10^9 M_\odot$. There is a clear gain in accuracy when including the VICS82 near-infrared photometry in the stellar mass estimates out to $z \approx 1$, and this is where the real benefit of the wide VICS82 survey can be found.

### 3.2. *Clusters: mass calibration, baryon census, and lensing*

Several optical cluster catalogs have already been constructed for Stripe 82 (e.g., Geach et al. 2011; Durret et al. 2015), and now sensitive millimetre mapping with the Atacama Cosmology Telescope (ACT) and *Planck* are producing *mass-limited* SZ-selected cluster samples (e.g. Hasselfield et al. 2013). For all clusters within the VICS82 footprint, we will be able to measure the total cluster mass via a stacked weak lensing technique (via the high quality CS82 *i*-band imaging, Shan et al. 2014), and the VICS82 data will enable us to measure the stellar mass function of cluster members down to $M_\star \approx 5 \times 10^{10} M_\odot$. For clusters detected by ACT, we can thus readily measure the baryon fraction in clusters to $z > 1$, which will be a strong constraint on cluster formation models. For lower mass clusters with a SZ signal below the ACT limit, we can stack the maps at the location of optically identified clusters to search for the average SZ signal, thus probing the baryon fraction to lower mass regimes. In Figure 10 we present examples of $iJK_s$ composite images of several $z > 0.5$ clusters identified by ACT through the SZ-effect (Hasselfield et al. 2013).

We can also search for strongly lensed background galaxies, revealing lensed galaxies that are extremely red in near-infrared/optical colours. These could be examples of high-$z$ dusty starburst galaxies, where the optical light is heavily extinguished, or massive and passive galaxies at $z > 1$. Strong lensing allows us to perform follow-up studies that would otherwise be impossible in the non-lensed case, owing to the flux amplification and magnification of projected scales by strong lensing. Geach et al. (2015) present a demonstration of the detection of a 'red arc', discovered as part of the citizen science project SPACEWARPS (Marshall et al. 2016; More et al. 2016). This project used 40,000 $iJK_s$ RGB composite images from VICS82 and CS82 data with the aim of identifying gravitationally lensed features. The best candidate was '9io9', a red partial Einstein ring around a Luminous Red Galaxy at $z \approx 0.2$. Subsequent spectroscopic (near-infrared and millimetre) follow-up determined the redshift of this source to be $z = 2.553$, and it was also revealed to be a radio- and sub-millimetre bright active galaxy of intrinsic luminosity $L > 10^{13} L_\odot$. Discovery of such rare sources is only made possible by large surveys such as VICS82 and we intend to mine the data for further discoveries, both through citizen science and in automatic machine learning searches (e.g., Hocking et al. 2015; Bom et al. 2016).

### 3.3. *Intermediate redshift quasars*

Observations of the quasar population in the near-infrared are key, since this links the rest-frame ultraviolet/optical to the mid-infrared ($\lambda \sim 5 - 30 \, \mu$m). However, only the bright, <16 magnitude quasars are detected in the shallow 2MASS survey; the majority of known quasars are fainter than this in the near-infrared bands. Peth et al. (2011) generated a catalog of 70,000 $K$-band detected QSOs over the SDSS DR6 footprint; those authors used UKIDSS LAS data on Stripe 82, but even with these slightly deeper data, the $i > 21$ mag objects were not detected. It was shown that using a KX selection (where the quasar SED shows an excess in the $K$-band compared to a stellar SED), one can successfully identify quasar candidate objects that would be normally excluded from the standard SDSS optical quasar selection algorithm (Peth et al. 2011). We demonstrate this for SDSS+VICS82 $gjK_s$ photometry in

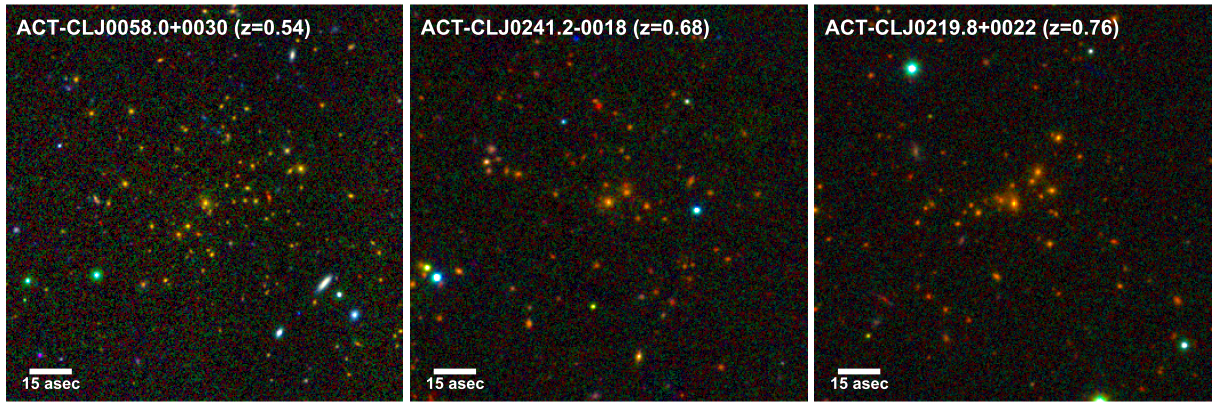

**Figure 10.** $iJK_s$ composite images of three distant clusters identified via the SZ-effect with ACT (Hasselfield et al. 2013). Each panel is $140''$ on a side. VICS82 can clearly identify the passive galaxy population of each cluster, which allows us not only to assess the stellar mass content of rich distant clusters, but also detect new clusters out to $z \approx 1$ with masses that place them below the limit of SZ surveys.

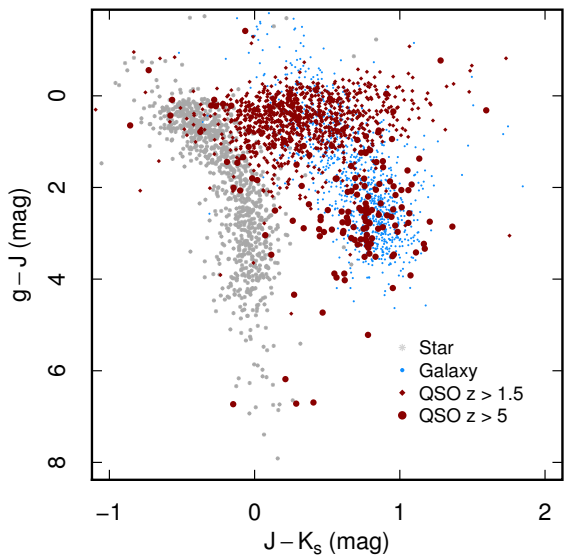

**Figure 11.** $gJK_s$ colour-colour plot showing spectroscopically classified objects from SDSS DR9 with $K_s < 21$ mag, where we separate out the very high redshift ($z > 5$) QSOs. For clarity we only show no more than 1000 of each type of object. This illustrates how intermediate and high redshift QSOs can be selected using a combination of SDSS optical and VICS82 near-infrared photometry, with extragalactic point sources cleanly separated from the stellar locus.

Figure 11, where we plot the colours of sources that have been spectroscopically identified as 'STAR', 'GALAXY' or 'QSO' by SDSS. Thus, VICS82 opens up the possibility of investigating the quasar epoch over $2 < z < 3.5$, where current (usually optically selected) quasar samples are poorly represented. The BOSS survey is probing these redshifts (e.g. Ross et al. 2012), but data from VICS82 is required to match the $i$-band depths of CS82 and DES.

## 4. SUMMARY

We present VICS82: the VISTA–CFHT near-infrared survey of Stripe 82. VICS82 comprises 150 square degrees of moderately deep ($J \approx 22$ mag, $K_s \approx 21.5$ mag) near-infrared imaging in what is becoming the first bona fide $\sim 100$ square degree scale extragalactic survey field. Around 9.5 million sources are catalogued down to the $K_s < 22$ mag, approximately 41% of which are matched to SDSS DR9 counterparts (including spectroscopy where available).

Naturally there exist a wide range of application for the VICS82 data, however in this paper we have outlined a few of the key goals that motivated the survey in the first instance. These include evaluating the stellar mass functions of $> L^\star$ galaxies out to $z \sim 1$, stellar mass calibration and baryon census of galaxy clusters, strong lensing, cross-correlation of optical/near-infrared selected galaxy catalogues with cosmic microwave background lensing, and the detection of intermediate redshift (KX-selected) quasars.

This article presents the VICS82 survey definition, including a description of the data acquisition and reduction methods, calibration, data quality analysis and source extraction. We make available the first VICS82 data release, comprising a catalogue of VICS82 $K_s$-band selected sources, matched to an independently extracted $J$-band catalogue. This is in turn matched to the SDSS photometric and spectroscopic catalogues where optical counterparts exist at SDSS Stripe 82 depths. Imaging (pixel) data is made available through a web tool. We now plan a series of data releases of increasing sophistication, culminating in a band-merged catalogue containing fully PSF-homogenised photometry across VICS82, SDSS, CS82, DES and SpIES, and including photometric redshifts and stellar mass estimates incorporating the new near-infrared data.

The catalogue and imaging are available from **http://stri-cluster.herts.ac.uk/vics82/**.

ACKNOWLEDGEMENTS

We thank the referee for a constructive report that has improved this paper. JEG thanks the Royal Society for a University Research Fellowship. YTL acknowledges support from the Ministry of Science and Technology grants MOST 104-2112-M-001-047 and 105-2112-M-001-028-MY3. MM is partially supported by CNPq (grant 312353/2015-4) and FAPERJ. GBC thanks the financial support from PRIN-INAF 2014 1.05.01.94.02.

This work is based on observations obtained with WIRCam, a joint project of Taiwan, Korea, Canada, France, and the Canada-France-Hawaii Telescope (CFHT) at CFHT, and VIRCAM at the VISTA/ESO Telescope at the Paranal Observatory under programme ID 090.A-0570. CFHT is operated by the National Research Council (NRC) of Canada, the Institut National des Sciences de l'Univers of the Centre National de la Recherche Scientifique (CNRS) of France, and the University of Hawaii. The Brazilian partnership on CFHT is managed by the Laboratório Nacional de Astrofísica (LNA). We thank Terapix and the Cambridge Astronomical Survey Unit for the VISTA data reduction. Access to the CFHT for the Taiwanese community was made possible by the contributions from Institute of Astronomy and Astrophysics, Academia Sinica.